%% file: main.tex
\documentclass[12pt,a4paper]{article}
\usepackage[utf8x]{inputenc}
\usepackage{longtable} 
\usepackage{rotating} 
\usepackage{booktabs}
\usepackage{float}
\usepackage{color}
\usepackage{epstopdf}
\usepackage{ifthen} 
\newboolean{pdflatex}
\setboolean{pdflatex}{true}
\newboolean{articletitles}
\setboolean{articletitles}{true} 
\newboolean{uprightparticles}
\setboolean{uprightparticles}{false} 
\newboolean{inbibliography}
\setboolean{inbibliography}{false}
\def\paperauthors{LHCb collaboration}
\def\paperasciititle{\  Search for the doubly heavy baryon Xibc via decays to DpK}
\def\papertitle{Search for 

the doubly heavy \Xibc baryon 

via decays to \Dz \proton \Km}
\def\paperkeywords{{High Energy Physics}, {LHCb}}
\def\papercopyright{\the\year\ CERN on behalf of the LHCb collaboration}
\def\paperlicence{CC BY 4.0 licence}
\def\paperlicenceurl{https://creativecommons.org/licenses/by/4.0/}
\input{preamble}
\input{xibc-symbols-def}

\usepackage{longtable}
\usepackage{multirow}

\begin{document}
\renewcommand{\thefootnote}{\fnsymbol{footnote}}
\setcounter{footnote}{1}
\input{title-LHCb-PAPER}

\renewcommand{\thefootnote}{\arabic{footnote}}
\setcounter{footnote}{0}
\pagestyle{plain}
\setcounter{page}{1}
\pagenumbering{arabic}
\input{xibc-paper-body}

\input{acknowledgements}

\addcontentsline{toc}{section}{References}
\setboolean{inbibliography}{true}
\bibliographystyle{LHCb}
\bibliography{standard,LHCb-PAPER,LHCb-CONF,LHCb-DP,LHCb-TDR,xibc-refs}
 
\newpage                                                                                                                                                                                                                                                                                                 
\newpage
\input{LHCb_Authorship_12-May-2020.tex}

\end{document}

%% file: preamble.tex
\usepackage[top=1in, bottom=1.25in, left=1in, right=1in]{geometry}

\usepackage{microtype}
\usepackage{lineno}  % for line numbering during review
\usepackage{xspace} % To avoid problems with missing or double spaces after
\usepackage{caption} %these three command get the figure and table captions automatically small

\usepackage{graphicx}  % to include figures (can also use other packages)
\usepackage{color}
\usepackage{colortbl}
\graphicspath{{./figs/}} % Make Latex search fig subdir for figures

\usepackage{amsmath} % Adds a large collection of math symbols
\usepackage{amssymb}
\usepackage{amsfonts}
\usepackage{upgreek} % Adds in support for greek letters in roman typeset

\newcommand*\patchAmsMathEnvironmentForLineno[1]{%
\expandafter\let\csname old#1\expandafter\endcsname\csname #1\endcsname
\expandafter\let\csname oldend#1\expandafter\endcsname\csname
end#1\endcsname
 \renewenvironment{#1}%
   {\linenomath\csname old#1\endcsname}%
   {\csname oldend#1\endcsname\endlinenomath}%
}
\newcommand*\patchBothAmsMathEnvironmentsForLineno[1]{%
  \patchAmsMathEnvironmentForLineno{#1}%
  \patchAmsMathEnvironmentForLineno{#1*}%
}
\AtBeginDocument{%
\patchBothAmsMathEnvironmentsForLineno{equation}%
\patchBothAmsMathEnvironmentsForLineno{align}%
\patchBothAmsMathEnvironmentsForLineno{flalign}%
\patchBothAmsMathEnvironmentsForLineno{alignat}%
\patchBothAmsMathEnvironmentsForLineno{gather}%
\patchBothAmsMathEnvironmentsForLineno{multline}%
\patchBothAmsMathEnvironmentsForLineno{eqnarray}%
}

\usepackage{hyperxmp}

\usepackage[pdftex,
            pdfauthor={\paperauthors},
            pdftitle={\paperasciititle},
            pdfkeywords={\paperkeywords},
            pdfcopyright={Copyright (C) \papercopyright},
            pdflicenseurl={\paperlicenceurl}]{hyperref}

\usepackage[all]{hypcap} % Internal hyperlinks to floats.

\input{lhcb-symbols-def} % Add in the predefined LHCb symbols

\usepackage{cite} % Allows for ranges in citations
\usepackage{mciteplus}

%% file: lhcb-symbols-def.tex
\usepackage{xspace} 
\usepackage{upgreek}

\def\lhcb {\mbox{LHCb}\xspace}

\def\lhc    {\mbox{LHC}\xspace}

\def\MagUp {\mbox{\em Mag\kern -0.05em Up}\xspace}

\ifthenelse{\boolean{uprightparticles}}%
{\def\Palpha      {\ensuremath{\upalpha}\xspace}

 \def\Pmu         {\ensuremath{\upmu}\xspace}                 
 \def\Pnu         {\ensuremath{\upnu}\xspace}                 
                  
 \def\Ppi         {\ensuremath{\uppi}\xspace}

 \def\Ppsi        {\ensuremath{\uppsi}\xspace}

 \def\PDelta      {\ensuremath{\Delta}\xspace}                 
 \def\PXi      {\ensuremath{\Xi}\xspace}                 
 \def\PLambda      {\ensuremath{\Lambda}\xspace}                 
 \def\PSigma      {\ensuremath{\Sigma}\xspace}                 
 \def\POmega      {\ensuremath{\Omega}\xspace}                 
 \def\PUpsilon      {\ensuremath{\Upsilon}\xspace}                 
 
 \def\PB      {\ensuremath{\mathrm{B}}\xspace}                 
                  
 \def\PD      {\ensuremath{\mathrm{D}}\xspace}

 \def\PJ      {\ensuremath{\mathrm{J}}\xspace}                 
 \def\PK      {\ensuremath{\mathrm{K}}\xspace}

 \def\PW      {\ensuremath{\mathrm{W}}\xspace}

 \def\Pb      {\ensuremath{\mathrm{b}}\xspace}                 
 \def\Pc      {\ensuremath{\mathrm{c}}\xspace}                 
 \def\Pd      {\ensuremath{\mathrm{d}}\xspace}

 \def\Pi      {\ensuremath{\mathrm{i}}\xspace}

 \def\Pp      {\ensuremath{\mathrm{p}}\xspace}

 \def\Ps      {\ensuremath{\mathrm{s}}\xspace}                 
                  
 \def\Pu      {\ensuremath{\mathrm{u}}\xspace}

}
{\def\Palpha      {\ensuremath{\alpha}\xspace}

 \def\Pmu         {\ensuremath{\mu}\xspace}                 
 \def\Pnu         {\ensuremath{\nu}\xspace}                 
                  
 \def\Ppi         {\ensuremath{\pi}\xspace}

 \def\Ppsi        {\ensuremath{\psi}\xspace}                 
                  
 \mathchardef\PDelta="7101
 \mathchardef\PXi="7104
 \mathchardef\PLambda="7103
 \mathchardef\PSigma="7106
 \mathchardef\POmega="710A
 \mathchardef\PUpsilon="7107
                  
 \def\PB      {\ensuremath{B}\xspace}                 
                  
 \def\PD      {\ensuremath{D}\xspace}

 \def\PJ      {\ensuremath{J}\xspace}                 
 \def\PK      {\ensuremath{K}\xspace}

 \def\PW      {\ensuremath{W}\xspace}

 \def\Pb      {\ensuremath{b}\xspace}                 
 \def\Pc      {\ensuremath{c}\xspace}                 
 \def\Pd      {\ensuremath{d}\xspace}

 \def\Pi      {\ensuremath{i}\xspace}

 \def\Pp      {\ensuremath{p}\xspace}

 \def\Ps      {\ensuremath{s}\xspace}                 
                  
 \def\Pu      {\ensuremath{u}\xspace}

}

\makeatletter
\ifcase \@ptsize \relax% 10pt
  \newcommand{\miniscule}{\@setfontsize\miniscule{4}{5}}% \tiny: 5/6
\or% 11pt
  \newcommand{\miniscule}{\@setfontsize\miniscule{5}{6}}% \tiny: 6/7
\or% 12pt
  \newcommand{\miniscule}{\@setfontsize\miniscule{5}{6}}% \tiny: 6/7
\fi
\makeatother

\DeclareRobustCommand{\optbar}[1]{\shortstack{{\miniscule (\rule[.5ex]{1.25em}{.18mm})}
  \\ [-.7ex] $#1$}}

\def\mup        {{\ensuremath{\Pmu^+}}\xspace}
\def\mun        {{\ensuremath{\Pmu^-}}\xspace} % muon negative (\mum is taken)

\def\ellm       {{\ensuremath{\ell^-}}\xspace}
\def\ellp       {{\ensuremath{\ell^+}}\xspace}
\def\ellell     {\ensuremath{\ell^+ \ell^-}\xspace}

\def\neu        {{\ensuremath{\Pnu}}\xspace}
\def\neub       {{\ensuremath{\overline{\Pnu}}}\xspace}
\def\neul       {{\ensuremath{\neu_\ell}}\xspace}
\def\neulb      {{\ensuremath{\neub_\ell}}\xspace}
\def\W      {{\ensuremath{\PW}}\xspace}
\def\Wp     {{\ensuremath{\PW^+}}\xspace}
\def\Wm     {{\ensuremath{\PW^-}}\xspace}

\def\uquark    {{\ensuremath{\Pu}}\xspace}

\def\dquark    {{\ensuremath{\Pd}}\xspace}

\def\squark    {{\ensuremath{\Ps}}\xspace}

\def\cquark    {{\ensuremath{\Pc}}\xspace}

\def\bquark    {{\ensuremath{\Pb}}\xspace}

\def\pion   {{\ensuremath{\Ppi}}\xspace}

\def\pip    {{\ensuremath{\pion^+}}\xspace}
\def\pim    {{\ensuremath{\pion^-}}\xspace}

\def\kaon    {{\ensuremath{\PK}}\xspace}

  \def\Kbar    {{\kern 0.2em\overline{\kern -0.2em \PK}{}}\xspace}

\def\KorKbar    {\kern 0.18em\optbar{\kern -0.18em K}{}\xspace}
\def\Kz      {{\ensuremath{\kaon^0}}\xspace}

\def\Km      {{\ensuremath{\kaon^-}}\xspace}

  \def\Dbar    {{\kern 0.2em\overline{\kern -0.2em \PD}{}}\xspace}
\def\D       {{\ensuremath{\PD}}\xspace}

\def\DorDbar    {\kern 0.18em\optbar{\kern -0.18em D}{}\xspace}
\def\Dz      {{\ensuremath{\D^0}}\xspace}
\def\Dzb     {{\ensuremath{\Dbar{}^0}}\xspace}

\def\Dstarp  {{\ensuremath{\D^{*+}}}\xspace}

\def\B       {{\ensuremath{\PB}}\xspace}
\def\Bbar    {{\ensuremath{\kern 0.18em\overline{\kern -0.18em \PB}{}}}\xspace}

\def\BorBbar    {\kern 0.18em\optbar{\kern -0.18em B}{}\xspace}
\def\Bz      {{\ensuremath{\B^0}}\xspace}

\def\Bcp     {{\ensuremath{\B_\cquark^+}}\xspace}

\def\jpsi     {{\ensuremath{{\PJ\mskip -3mu/\mskip -2mu\Ppsi\mskip 2mu}}}\xspace}

  \def\Y#1S{\ensuremath{\PUpsilon{(#1S)}}\xspace}% no space before {...}!

\def\proton      {{\ensuremath{\Pp}}\xspace}

\def\Xires       {{\ensuremath{\PXi}}\xspace}

\def\Lz          {{\ensuremath{\PLambda}}\xspace}
\def\Lbar        {{\ensuremath{\kern 0.1em\overline{\kern -0.1em\PLambda}}}\xspace}
\def\LorLbar    {\kern 0.18em\optbar{\kern -0.18em \PLambda}{}\xspace}

\def\Lb      {{\ensuremath{\Lz^0_\bquark}}\xspace}

\def\Xibz    {{\ensuremath{\Xires^0_\bquark}}\xspace}

\def\BF         {{\ensuremath{\mathcal{B}}}\xspace}

\def\BR         {\BF}
\def\to                 {\ensuremath{\rightarrow}\xspace}

\def\AT#1     {\ensuremath{A_{\mathrm{T}}^{#1}}\xspace}           % 2

\def\C#1      {\ensuremath{\mathcal{C}_{#1}}\xspace}                       % 9
\def\Cp#1     {\ensuremath{\mathcal{C}_{#1}^{'}}\xspace}                    % 7
\def\Ceff#1   {\ensuremath{\mathcal{C}_{#1}^{\mathrm{(eff)}}}\xspace}        % 9  
\def\Cpeff#1  {\ensuremath{\mathcal{C}_{#1}^{'\mathrm{(eff)}}}\xspace}       % 7
\def\Ope#1    {\ensuremath{\mathcal{O}_{#1}}\xspace}                       % 2
\def\Opep#1   {\ensuremath{\mathcal{O}_{#1}^{'}}\xspace}                    % 7

\newcommand{\tev}{\ifthenelse{\boolean{inbibliography}}{\ensuremath{~T\kern -0.05em eV}}{\ensuremath{\mathrm{\,Te\kern -0.1em V}}}\xspace}
\newcommand{\gev}{\ensuremath{\mathrm{\,Ge\kern -0.1em V}}\xspace}
\newcommand{\mev}{\ensuremath{\mathrm{\,Me\kern -0.1em V}}\xspace}
\newcommand{\kev}{\ensuremath{\mathrm{\,ke\kern -0.1em V}}\xspace}
\newcommand{\ev}{\ensuremath{\mathrm{\,e\kern -0.1em V}}\xspace}
\newcommand{\gevc}{\ensuremath{{\mathrm{\,Ge\kern -0.1em V\!/}c}}\xspace}
\newcommand{\mevc}{\ensuremath{{\mathrm{\,Me\kern -0.1em V\!/}c}}\xspace}
\newcommand{\gevcc}{\ensuremath{{\mathrm{\,Ge\kern -0.1em V\!/}c^2}}\xspace}
\newcommand{\gevgevcccc}{\ensuremath{{\mathrm{\,Ge\kern -0.1em V^2\!/}c^4}}\xspace}
\newcommand{\mevcc}{\ensuremath{{\mathrm{\,Me\kern -0.1em V\!/}c^2}}\xspace}

\def\mum  {\ensuremath{{\,\upmu\mathrm{m}}}\xspace}

\def\nb {\ensuremath{\mathrm{ \,nb}}\xspace}

\def\invfb   {\ensuremath{\mbox{\,fb}^{-1}}\xspace}

\def\fs   {\ensuremath{\mathrm{ \,fs}}\xspace}

\newcommand{\chisq}{\ensuremath{\chi^2}\xspace}

\newcommand{\chisqip}{\ensuremath{\chi^2_{\text{IP}}}\xspace}

\def\gsim{{~\raise.15em\hbox{$>$}\kern-.85em
          \lower.35em\hbox{$\sim$}~}\xspace}
\def\lsim{{~\raise.15em\hbox{$<$}\kern-.85em
          \lower.35em\hbox{$\sim$}~}\xspace}

\def\sqs   {\ensuremath{\protect\sqrt{s}}\xspace}

\def\ptot       {\mbox{$p$}\xspace}
\def\pt         {\mbox{$p_{\mathrm{ T}}$}\xspace}

\def\evtgen     {\mbox{\textsc{EvtGen}}\xspace}

\def\geant      {\mbox{\textsc{Geant4}}\xspace}

\def\photos     {\mbox{\textsc{Photos}}\xspace}

\def\pythia     {\mbox{\textsc{Pythia}}\xspace}

\def\tell1  {TELL1\xspace}
\def\ukl1   {UKL1\xspace}

\newcommand{\eg}{\mbox{\itshape e.g.}\xspace}
\newcommand{\ie}{\mbox{\itshape i.e.}\xspace}

%% file: xibc-symbols-def.tex
\usepackage{xspace} 
\usepackage{upgreek}

\def\DeltaOrDeltabar  {\kern 0.18em\optbar{\kern -0.18em \PDelta}{}\xspace}
\def\XiOrXibar        {\kern 0.18em\optbar{\kern -0.18em \PXi}{}\xspace}
\def\SigmaOrSigmabar  {\kern 0.18em\optbar{\kern -0.18em \PSigma}{}\xspace}
\def\OmegaOrOmegabar  {\kern 0.18em\optbar{\kern -0.18em \POmega}{}\xspace}

\def\Xiccpp       {{\ensuremath{\Xires^{++}_{\cquark\cquark}}}\xspace}

\newcommand{\TeVnosp}{\ifthenelse{\boolean{inbibliography}}{\ensuremath{~T\kern -0.05em eV}}{\ensuremath{\mathrm{Te\kern -0.1em V}}}}
\newcommand{\GeVnosp}{\ensuremath{\mathrm{Ge\kern -0.1em V}}}
\newcommand{\MeVnosp}{\ensuremath{\mathrm{Me\kern -0.1em V}}}
\newcommand{\keVnosp}{\ensuremath{\mathrm{ke\kern -0.1em V}}}
\newcommand{\eVnosp}{\ensuremath{\mathrm{e\kern -0.1em V}}}
\newcommand{\GeVcnosp}{\ensuremath{{\mathrm{Ge\kern -0.1em V\!/}c}}}
\newcommand{\MeVcnosp}{\ensuremath{{\mathrm{Me\kern -0.1em V\!/}c}}}
\newcommand{\GeVccnosp}{\ensuremath{{\mathrm{Ge\kern -0.1em V\!/}c^2}}}
\newcommand{\MeVccnosp}{\ensuremath{{\mathrm{Me\kern -0.1em V\!/}c^2}}}
\newcommand{\GeVGeVccccnosp}{\ensuremath{{\mathrm{Ge\kern -0.1em V^2\!/}c^4}}}

\def\genxicctwo   {\mbox{\textsc{GenXicc2.0}}\xspace}

\def\Xiccpp{{\ensuremath{{\Xires}_{\cquark\cquark}^{++}}}\xspace}

\def\Xibc{{\ensuremath{{\Xires}_{\bquark\cquark}^{0}}}\xspace}
\def\Xibcp{{\ensuremath{{\Xires}_{\bquark\cquark}^{+}}}\xspace}
\def\XibcDecay{{\ensuremath{\Xibc\to\Dz \proton \Km}}\xspace}
\def\XibDecay{{\ensuremath{\Xibz\to\Dz \proton \Km}}\xspace}
\def\XibcpDecay{{\ensuremath{\Xibcp\to\Dz \proton \Kz}}\xspace}
\def\LbDecay{{\ensuremath{\Lb\to\Dz \proton \Km}}\xspace}
\def\LbDecaymis{{\ensuremath{\Lb\to\Dz \proton \pim}}\xspace}

\def\DzDecay{{\ensuremath{\Dz\to \Km \pip}}\xspace}
\def\BDecay{{\ensuremath{\Bz\to\Dzb \pip \pim}}\xspace}

\def\nsig{\ensuremath{N_{\mathrm{sig}}}\xspace}
\def\ncon{\ensuremath{N_{\mathrm{norm}}}\xspace}
\def\effSig{\ensuremath{\epsilon_{\mathrm{sig}}}\xspace}
\def\effCon{\ensuremath{\epsilon_{\mathrm{norm}}}\xspace}

%% file: title-LHCb-PAPER.tex
\begin{titlepage}
\pagenumbering{roman}

\vspace*{-1.5cm}
\centerline{\large EUROPEAN ORGANIZATION FOR NUCLEAR RESEARCH (CERN)}
\vspace*{1.5cm}
\noindent
\begin{tabular*}{\linewidth}{lc@{\extracolsep{\fill}}r@{\extracolsep{0pt}}}
\ifthenelse{\boolean{pdflatex}}
{\vspace*{-2.7cm}\mbox{\!\!\!\includegraphics[width=.14\textwidth]{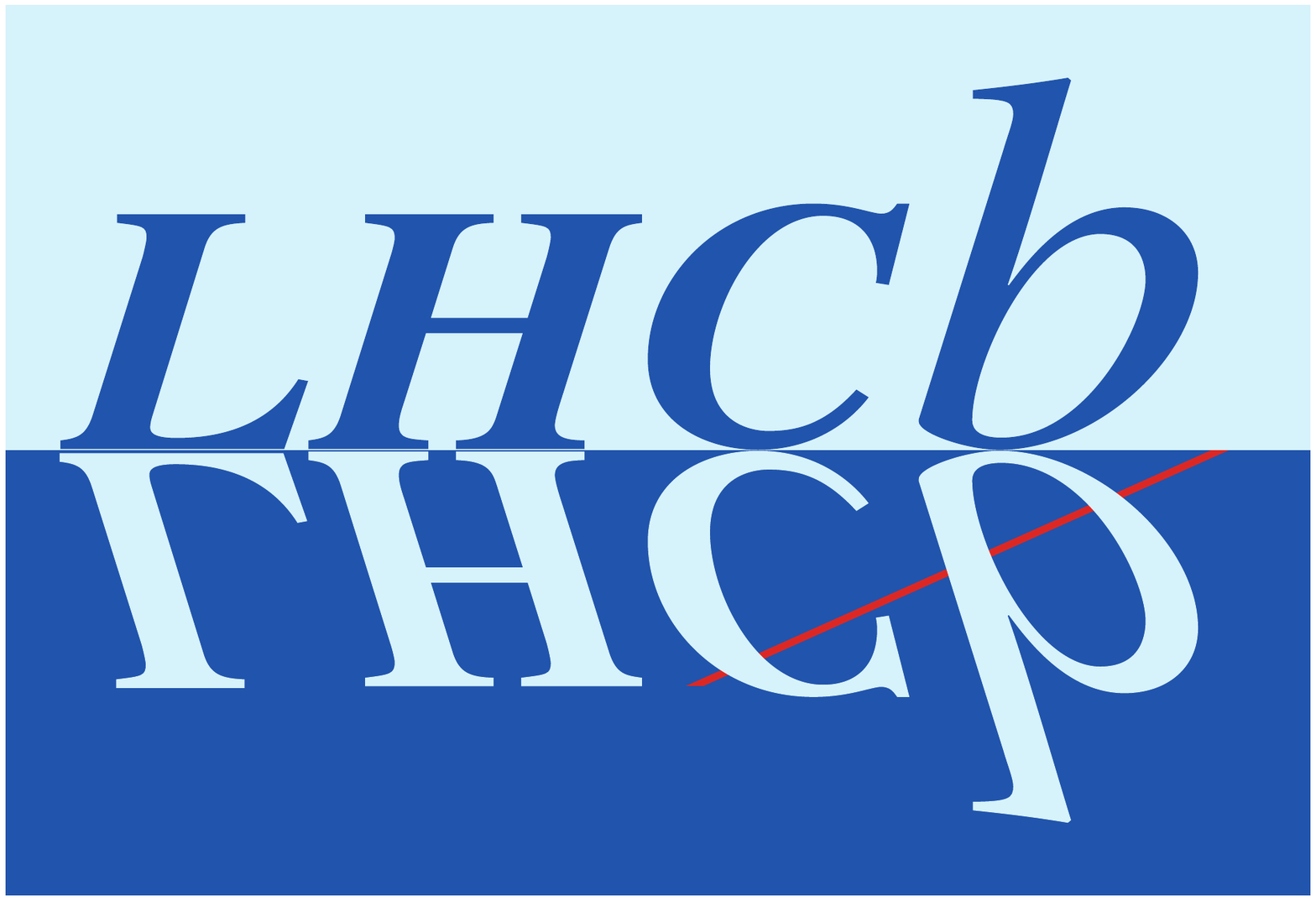}} & &}
{\vspace*{-1.2cm}\mbox{\!\!\!\includegraphics[width=.12\textwidth]{lhcb-logo.eps}} & &}
\\
 & & CERN-EP-2020-137 \\  
 & & LHCb-PAPER-2020-014 \\  
 & & November 23, 2020 \\ 
\end{tabular*}

\vspace*{4.0cm}

{\normalfont\bfseries\boldmath\huge
\begin{center}

  \papertitle 
\end{center}
}

\vspace*{2.0cm}

\begin{center}

\paperauthors\footnote{Authors are listed at the end of this paper.}
\end{center}

\vspace{\fill}

\begin{abstract}
  \noindent
  A search for the doubly heavy \Xibc baryon using its decay to the $D^0pK^-$ final state
  is performed using proton-proton collision data at a centre-of-mass
  energy of 13\tev collected by the LHCb experiment between 2016 and 2018,
  corresponding to an integrated luminosity of 5.4\invfb. No
  significant signal is found in the invariant mass range from 6.7 to 7.2 \gevcc. 
  Upper limits are set at 95\% credibility level on the ratio of the \Xibc
  production cross-section times its branching fraction to $D^0pK^-$
  relative to that of the \LbDecay decay. The limits are set as a function
  of the $\Xibc$ mass and lifetime hypotheses, in the rapidity range
  from 2.0 to 4.5 and in the transverse momentum region from 5 to 25
  \gevc. Upper limits range from $1.7\times10^{-2}$ to $3.0\times10^{-1}$ for the considered \Xibc mass and lifetime hypotheses.
  
\end{abstract}

\vspace*{2.0cm}

\begin{center}
  Published in JHEP
\end{center}

\vspace{\fill}

{\footnotesize 
\centerline{\copyright~\papercopyright, license \href{\paperlicenceurl}{\paperlicence}.}}
\vspace*{2mm}

\end{titlepage}
\newpage
\setcounter{page}{2}
\mbox{~}

\cleardoublepage

%% file: xibc-paper-body.tex
\section{Introduction}
\label{sec:Introduction}
In the constituent quark model~\cite{
  GellMann:1964nj,
  Zweig:352337,
  *Zweig:570209}, two heavy quarks (\bquark or \cquark) can
be bound together with a light quark to form doubly heavy baryons~\cite{PhysRevD.38.1519}. Studies of these particles are
of great interest for the understanding of  hadron spectroscopy and QCD at low energies. The \Xiccpp baryon (valence quark content
\cquark\cquark\uquark)\footnote{The inclusion of charge-conjugate modes is implied throughout this paper.} was first observed in 2017 by the \lhcb
collaboration~\cite{LHCb-PAPER-2017-018}. The \Xibc baryon (\bquark\cquark\dquark) containing two different heavy quarks
is expected to have a mass in the range of
$6.8-7.1\gevcc$~\cite{Ebert:2002ig,Gershtein:2000nx,Ebert:1996ec,Roncaglia:1994ex,NARODETSKII_2003,Roncaglia:1995az,Lichtenberg:1995kg,Kiselev:2001fw,He:2004px,Albertus:2006ya,Roberts:2007ni,Gerasyuta:2008zy,Weng:2010rb,Zhang:2008rt,Karliner:2014gca,Aliev:2012ru,Narison:2010py}.
The \Xibc production cross-section is predicted to be about 16\nb at a centre-of-mass energy of
$\sqs=13\tev$ in the pseudorapidity range $1.9<\eta <4.9$
and for a transverse momentum $\pt>4\gevc$~\cite{Zhang:2011hi}.

The \Xibc baryon has not been observed to date. Five categories of \Xibc decays have been studied theoretically: (i) semileptonic decays
induced by
${\ensuremath{\cquark\to\dquark(\squark)\ellp\neul}}\xspace$ or
${\ensuremath{\bquark\to\uquark(\cquark)\ellm\neulb}}\xspace$ transitions,
with branching fractions estimated to
be within the range
$10^{-6}-10^{-2}$~\cite{Hu:2020mxk,Shi:2019hbf,Wang:2017mqp,Gerasimov:2019jwp,Zhao:2018mrg,Berezhnoy:2018bde};
(ii) non-leptonic decays mediated by weak scattering of the \bquark-quark and \cquark-quark
~\cite{Kiselev:2001fw,Li:2017ndo};
(iii) non-leptonic decays occurring through \cquark-quark charged current interaction, whose
branching fractions are predicted to be
$10^{-5}-10^{-1}$~\cite{Shi:2019hbf,Wang:2017mqp,Gerasimov:2019jwp,Zhao:2018mrg,Berezhnoy:2018bde};
(iv) non-leptonic decays produced by \bquark-quark charged current,
with branching fractions ranging
$10^{-9}-10^{-3}$~\cite{Shi:2019hbf,Wang:2017mqp,Gerasimov:2019jwp,Zhao:2018mrg,Kiselev:2003dk}; and
(v) flavour-changing neutral current processes
${\ensuremath{\bquark\to\dquark(\squark)\ellell}}\xspace$, with
branching fractions highly suppressed and within the range
$10^{-10}-10^{-8}$~\cite{Hu:2020mxk,Xing:2018lre}.

The $\Xibc$ lifetime is estimated by calculating full decay width which is expected to consist of four major contributions, due to
{\ensuremath{\bquark\to\cquark\Wm}}\xspace and
{\ensuremath{\cquark\to\squark\Wp}}\xspace transitions, Pauli interference between the products of heavy quark decays and the quarks in the initial
state, and weak
scattering effects between the constituents, \eg \
{\ensuremath{\bquark\cquark\to\cquark\squark}}\xspace,
{\ensuremath{\cquark\dquark\to\squark\uquark}}\xspace. The \Xibc lifetime is predicted to be in the range of
$90-280\fs$~\cite{Karliner:2014gca,Cheng:2019sxr,Kiselev:2001fw,Kiselev:1999kh,Bjorken:1986kfa}.
By contrast, Ref.~\cite{SanchisLozano:1994vh} advocates that the $\Xibc$ lifetime is similar to that of the $\Bcp$ meson, \ie $(510 \pm 9)\fs$~\cite{PDG2019}.

\begin{figure}
  \centering
  \includegraphics[width=0.6\linewidth]{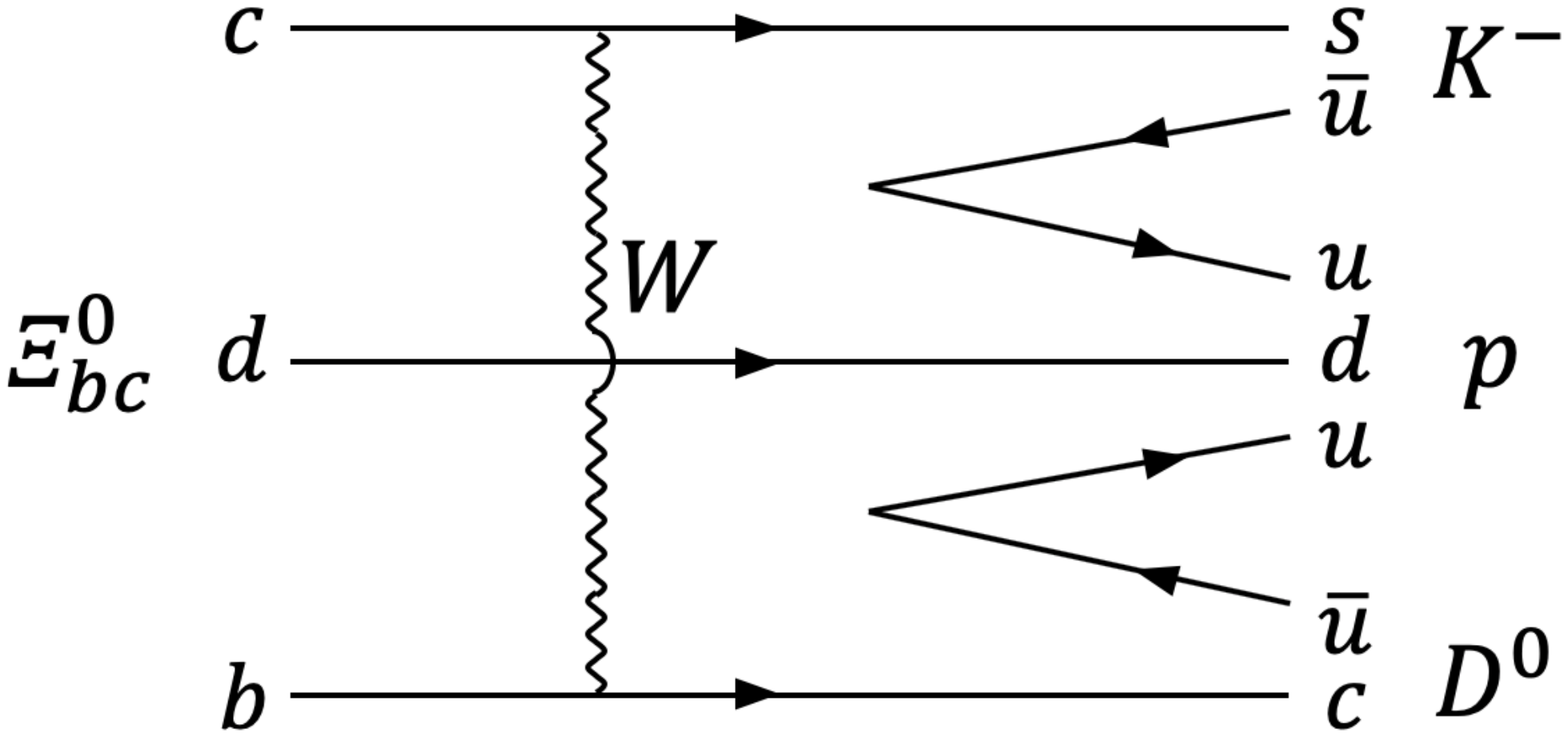}
  \caption{ The \XibcDecay decay induced by the weak \W-scattering of constituent \bquark and \cquark quarks.}
  \label{fig:diagram}
\end{figure}

This paper presents the first search for the \Xibc baryon in the
mass range from 6.7 to $7.2\gevcc$, using proton-proton ($pp$) collision data collected by
the LHCb experiment at a centre-of-mass energy of $\sqs=13 \tev$ between 2016 and 2018, corresponding to an integrated luminosity of 5.4\invfb.
The \Xibc baryon is searched for through the \XibcDecay, \DzDecay decay chain, which is preferred for its ease of reconstruction at \lhcb. A leading-order Feynman diagram contributing to this decay is shown in
Fig.~\ref{fig:diagram}.
The branching fraction $\BF(\XibcDecay)$ is
expected to be similar to that of the \XibcpDecay decay,
about 0.1$\%$~\cite{Kiselev:2001fw}.
Considering the value of $\BR(\DzDecay)=(3.89\pm 0.04)\%$~\cite{PDG2019},
the total branching fraction of 
the $\Xibc \to \Dz(\to \Km \pip)\proton \Km$ decay chain is expected to be in the range of $10^{-5}-10^{-4}$.

To reduce systematic uncertainties, the \Xibc production cross-section
is measured relative to that of the normalisation mode corresponding to a \Lb baryon decaying to the same final
state. Both the \Xibc and \Lb baryons are reconstructed in the rapidity range from 2.0 to 4.5 and in the transverse momentum region from 5 to 25\gevc.
The search is performed with the analysis procedure entirely defined before inspecting the data across the considered mass range.

\section{Detector and simulation}
\label{sec:Detector}
The \lhcb detector~\cite{LHCb-DP-2008-001,LHCb-DP-2014-002} is a single-arm forward
spectrometer covering the \mbox{pseudorapidity} range $2<\eta <5$,
designed for the study of particles containing \bquark or \cquark
quarks. The detector includes a high-precision tracking system
consisting of a silicon-strip vertex detector surrounding the $pp$
interaction region~\cite{LHCb-DP-2014-001}, a large-area silicon-strip detector located
upstream of a dipole magnet with a bending power of about
$4{\mathrm{\,Tm}}$, and three stations of silicon-strip detectors and straw
drift tubes~\cite{LHCb-DP-2017-001} placed downstream of the magnet. 
The polarity of the dipole magnet is reversed periodically throughout data taking.
The tracking system provides a measurement of the momentum, \ptot, of charged particles with
a relative uncertainty that varies from 0.5\% at low momentum to 1.0\% at 200\gevc.
The minimum distance of a track to a primary $pp$ interaction vertex (PV), the impact parameter (IP),
is measured with a resolution of $(15+29/\pt)\mum$, where $\pt$ is expressed in $\gevc$.
Different types of charged hadrons are distinguished using information
from two ring-imaging Cherenkov detectors~\cite{LHCb-DP-2012-003}.
Photons, electrons and hadrons are identified by a calorimeter system consisting of
scintillating-pad and preshower detectors, an electromagnetic and a hadronic calorimeter. Muons are identified by a
system composed of alternating layers of iron and multiwire
proportional chambers~\cite{LHCb-DP-2012-002}.
The online event selection is performed by a trigger~\cite{LHCb-DP-2012-004}, 
which consists of a hardware stage, based on information from the calorimeters and muon
systems~\cite{LHCb-DP-2013-001}, followed by a
software stage,  at which all tracks with $\pt>300 \mevc$ are
reconstructed for data collected at
$\sqs=13\tev$~\cite{LHCb-DP-2019-001}.
The software trigger used in this analysis requires a two-, three- or
four-track vertex with significant displacement from any
PV.
At least one charged particle must have $\pt>1.7\gevc$ and be inconsistent with originating from any PV. 
A multivariate algorithm~\cite{Gligorov:2012qt} is used for the
identification of displaced vertices consistent with the decay of a
\bquark hadron.

Simulated samples are used to develop the candidate selection and
to estimate the corresponding efficiency as well as that of the detector acceptance.
Simulated $pp$ collisions are generated using
\pythia~\cite{Sjostrand:2007gs,*Sjostrand:2006za} 
with a specific \lhcb configuration~\cite{LHCb-PROC-2010-056}.  
A dedicated package, \genxicctwo~\cite{Chang:2009va},
is used to simulate the \Xibc baryon production.
Decays of unstable particles
are described by \evtgen~\cite{Lange:2001uf}, in which final-state
radiation is generated using \photos~\cite{Golonka:2005pn}. The
interaction of the generated particles with the detector, and its response,
are simulated using the \geant
toolkit~\cite{Allison:2006ve, *Agostinelli:2002hh} as described in
Ref.~\cite{LHCb-PROC-2011-006}. The simulated $\Xibc$ events are generated with a mass of 6.9\gevcc and a lifetime of $400\fs$, and samples with different mass and lifetime hypotheses are obtained using a weighting technique. 
The $\Xibc$ baryon decay is assumed to follow a uniform phase-space model. 

\section{Reconstruction and selection}
\label{sec:Sel}
For both the \Xibc signal and the \Lb normalisation modes, \Dz candidates are
reconstructed in the \Km\pip final state.
Two oppositely charged tracks identified as a kaon and a pion with an invariant mass in the range of $1.84 < m(\Km\pip) <
1.89\gevcc$ 
are requested to form a 
common vertex that is significantly displaced from any PV. 
The \Dz candidate is then combined with two oppositely charged tracks identified as a proton and as a kaon to form a $\Xibc$ or a $\Lb$ candidate.
The two tracks are required to have a high transverse
momentum and to be inconsistent with originating from any PV. The \Dz, \proton and \kaon candidates are required to form a common vertex with a good fit quality. The $\Xibc$ and $\Lb$ candidates have to point back to the PV
and have an invariant mass larger than 5.0\gevcc.

A multivariate analysis is applied to both the signal and the normalisation candidates to further improve the purity of the samples. The selection algorithm is a Boosted Decision Tree (BDT) algorithm implemented in the TMVA package~\cite{Hocker:2007ht}. To train this classifier, simulated \Xibc baryon decays are used as the signal proxy and candidates lying in the upper \Dz\proton\Km mass sideband ($8.0-8.5\gevcc$) adjacent to the signal region for the background proxy. The BDT algorithm uses kinematic and vertex-topology variables that show good discrimination power between signal and background. 
The variables include: 
the $\chisqip$ and 
transverse momentum of all 
particles; particle identification (PID) variables for the final state particles; the flight-distance $\chisq$ between the PV and the decay vertex; the vertex quality of the $\Dz$ and $\Xibc$ candidates; and the
angle between the momentum and the flight direction of
the $\Xibc$ candidate. 
The $\chisqip$ is defined as the difference in $\chisq$ of the PV fit with and without the particle in question. The flight-distance $\chisq$ is defined as the $\chisq$ of
the hypothesis that the decay vertex of the candidate coincides with its associated PV, defined as the PV
with the smallest $\chisqip$. It has been verified that this BDT classifier does not shape the background invariant mass distribution.

A selection requirement is applied on the BDT response. It is determined by maximizing the value of the Punzi figure of merit $\varepsilon/(\frac{a}{2}+\sqrt{N_{B}})$ ~\cite{Punzi:2003bu}, where $\varepsilon$ is the estimated signal efficiency, $a$ corresponds to the number of standard deviations in a Gaussian significance test, which is taken as 5, and $N_{B}$ is the number of background candidates determined in the upper sideband and extrapolated to the signal region. 
The performance of the BDT classifier is tested and found to be stable against the $\Xibc$ lifetime in the range from 100 to 500$\fs$. 

\section{Yield measurements}
\label{sec:signals}
\begin{figure}
  \centering
  \includegraphics[width=0.75\linewidth]{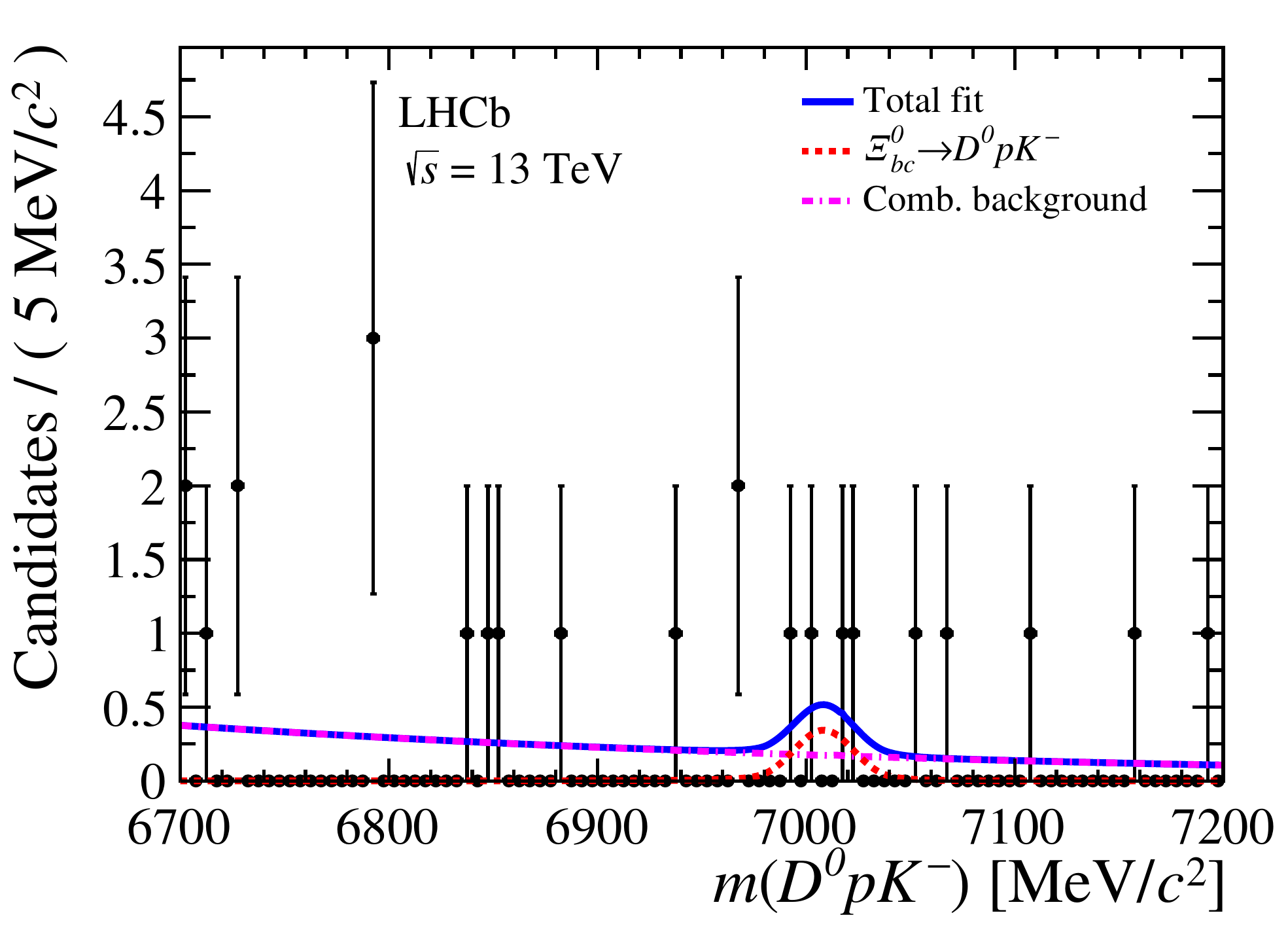}
  \caption{Invariant mass $m(\Dz\proton\Km)$ distribution of selected \Xibc candidates (black points) together with the projection of the fit (blue solid line) for the full data sample. The \XibcDecay signal component, with the central mass value varying freely (red dashed line), and combinatorial background (purple dotted line) are also shown.}
  \label{fig:dataset}
\end{figure}

The invariant mass distribution of the selected candidates within the range $6.7-7.2\gevcc$ for the full data sample is shown in Fig.~\ref{fig:dataset}. 
The \Xibc signal yield is determined from an unbinned maximum-likelihood fit to the invariant mass $m(\Dz\proton\Km)$ distribution.
The signal is described by a double-sided Crystal Ball (DSCB) function~\cite{Skwarnicki:1986xj} comprising a Gaussian core with power-law tails on both sides, while the background is described by an exponential function.
The parameters of the signal model are fixed from simulation except for the peak position that is allowed to vary in the fit. The mass resolution of the signal decay is $14.2\pm0.4\mevcc$ for all mass hypotheses, as determined from simulation. 
The projection of the fit to the mass distribution, with the \Xibc mass parameter varying freely, is also shown in Fig.~\ref{fig:dataset}.
No excess is observed in the full \Xibc mass range, therefore upper limits are set on the production ratios. 

\begin{figure}
  \centering
  \includegraphics[width=0.75\linewidth]{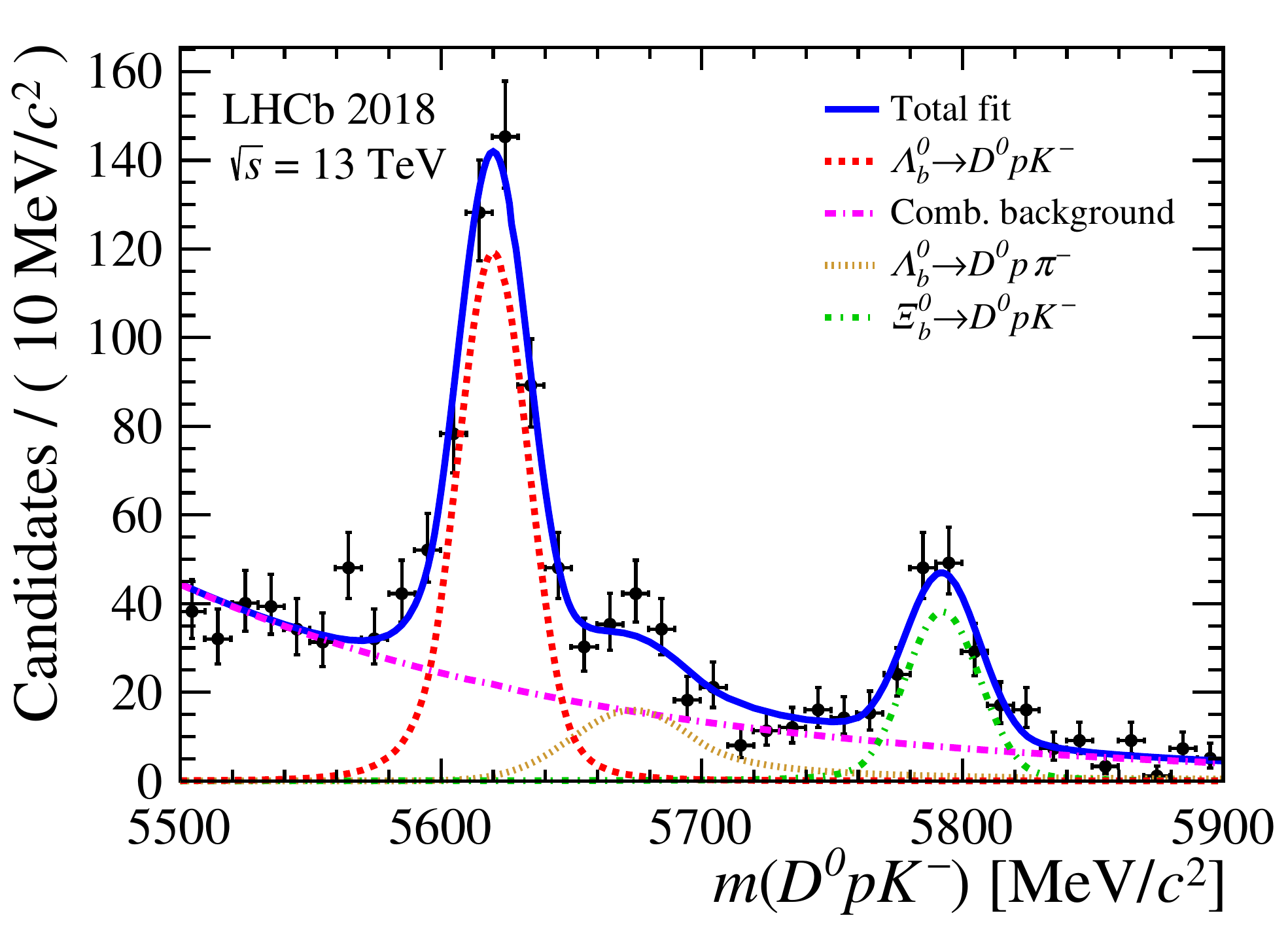}
  \caption{Invariant mass distribution for \LbDecay candidates in the 2018 data sample (black points). The fit projection (blue solid line) is superimposed. The normalisation component (red dashed line), the misidentified background (brown dashed line), the combinatorial background (purple dotted line), and the \XibDecay (green dotted line) components are also shown. Similar distributions are obtained for the 2016 and 2017 data samples.}
  \label{fig:lb_yield_fit_18}
\end{figure}

As the selection efficiency varies with the data-taking conditions, the yield of the normalisation mode is determined for each year separately.  
The \Lb signal yield, \ncon, is obtained from an extended unbinned maximum-likelihood fit to the invariant mass $m(\Dz\proton\Km)$ distribution in the 2016, 2017 and 2018 data samples. The fit model includes a DSCB function to describe the $\mbox{\LbDecay}$ decay and three separate background components: random combinations of tracks or genuine \Dz decays combined with random tracks (combinatorial background); the Cabibbo-favoured decay \LbDecaymis where the pion is incorrectly identified as a kaon (misidentified background); and the $\mbox{\XibDecay}$ decay component. The shape of the normalisation mode and the misidentified background are taken from simulation. The latter is parameterised with a Crystal Ball (CB) function. The \XibDecay decay component is described by a DSCB function and the combinatorial background by an exponential function. 
As an illustration, the $m(\Dz\proton\Km)$ distribution for the 2018 data sample is shown in Fig.~\ref{fig:lb_yield_fit_18} along with the projection of the associated fit result. A total of about 1200 \Lb candidates are obtained.

\section{Production cross-section ratio}
\label{sec:eff}

The production cross-section ratio, $R$, is defined as
\begin{equation}
R\equiv \frac{\sigma(\Xibc)\BR(\XibcDecay)}{\sigma(\Lb)\BR(\LbDecay)}
= \frac{\effCon}{\effSig}\frac{\nsig}{\ncon} \equiv\alpha\nsig,
\label{eq:ratiodefination}
\end{equation}
where $\sigma$ is the production cross-section and \BR is the decay branching fraction, $\effSig$ and $\effCon$ are the selection efficiencies
of the signal and normalisation decay modes, respectively, $\nsig$ and $\ncon$ are the corresponding yields, and $\alpha=\effCon/(\effSig\ncon)$ is the single-event sensitivity.

The signal efficiency depends upon the assumed mass and lifetime of the \Xibc. Simulated events are generated with $m(\Xibc)=6.9\gevcc$ and $\tau(\Xibc)=400\fs$, from here on referred to as nominal, and used to evaluate the efficiency ratio. The variation of the efficiency ratio as a function of $m(\Xibc)$ and $\tau(\Xibc)$ relative to the nominal point is then determined with a weighting technique discussed in Sec.~\ref{sec:effmtau}. The kinematic distribution of \Xibc baryons produced at the \lhc is also unknown and is assumed to be the same as for the \Lb baryon. Transverse momentum and rapidity distributions of simulated \Xibc are therefore corrected to match that of \Lb decays observed in data.

The efficiencies can be factorised into that of the geometrical acceptance, track reconstruction, trigger, offline pre-selection, PID, and multivariate selection. 
The individual efficiencies are evaluated with simulated events of \XibcDecay and \LbDecay decays, except for tracking and PID where the efficiencies are determined using calibration data samples, 
namely the $\mbox{\jpsi\to\mup\mun}$ decay~\cite{LHCb-DP-2013-002} for tracking and \Dstarp\to\Dz(\to\Km\pip)\pip and \Lz\to\proton\pim decays for PID~\cite{LHCb-PUB-2016-021,Aaij:2018vrk}.

The track multiplicity distribution is taken from \LbDecay data for both signal and normalisation samples. The simulated Dalitz plot of these decays are corrected to match the distribution observed in background-subtracted data, obtained using the {\it sPlot} technique~\cite{Pivk:2004ty}.
The efficiency ratio and the single-event sensitivity
at the nominal $\Xibc$ mass and lifetime are summarised in Table~\ref{tab:alpha}. 
The single-event sensitivity is determined according to Eq.~(\ref{eq:ratiodefination}) using the obtained efficiency ratios and the normalisation yields reported in Table~\ref{tab:alpha}.

\begin{figure}[b]
  \centering
  \includegraphics[width=0.88\linewidth]{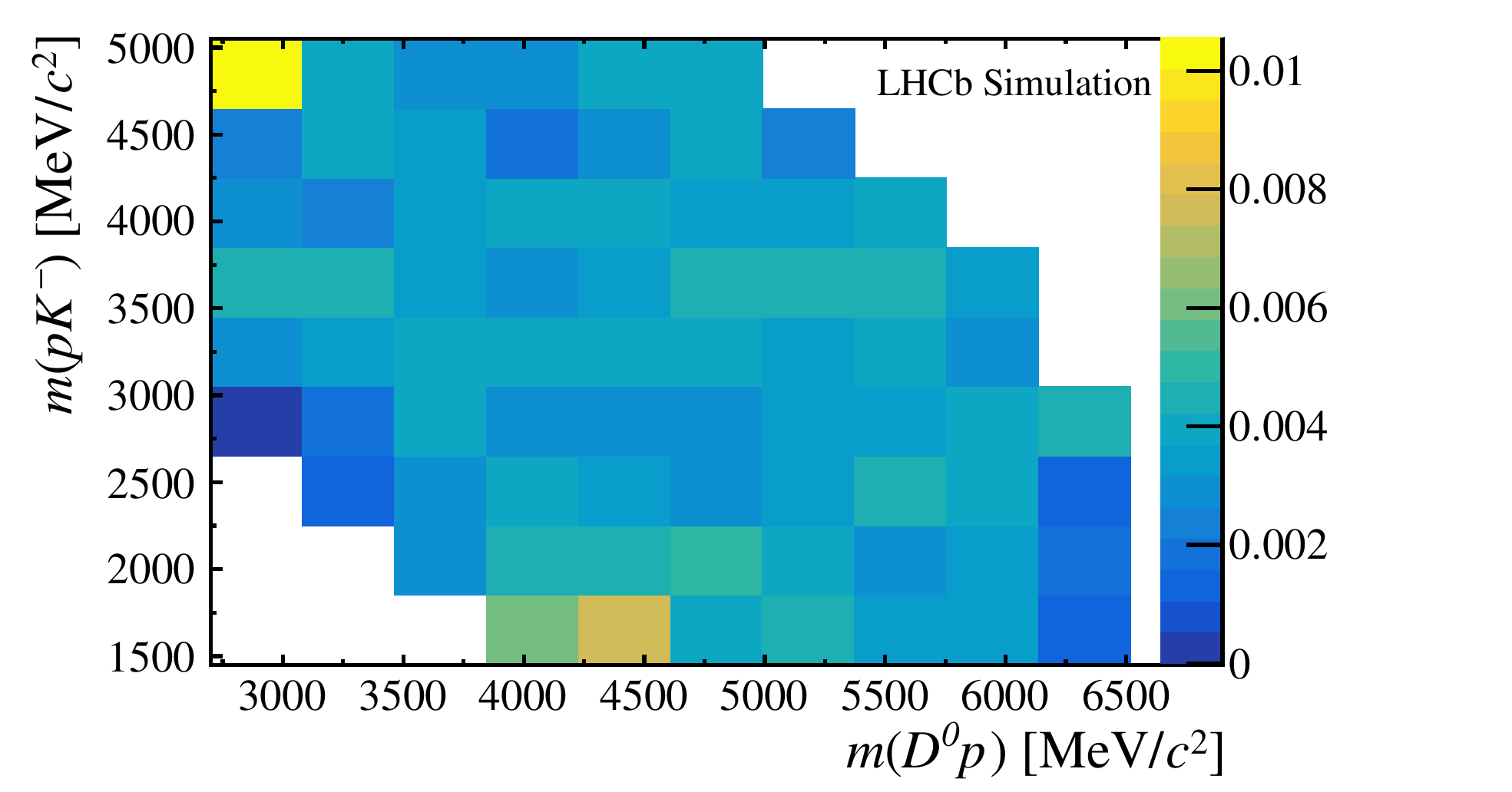}
  \caption{Efficiency of selected \XibcDecay decays as a function of the invariant masses $m(\Dz\proton)$ and $m(\proton\Km)$ in the simulation. The variation of efficiency across the Dalitz plot reflects the specific phase-space dependent requirements of the selection.}
  \label{fig:eff_vs_dalitz}
\end{figure}

The analysis is performed assuming a uniform phase-space model for the signal decay $\mbox{\XibcDecay}$. Efficiency maps in bins of the invariant masses $m(\Dz\proton)$ and $m(\proton\Km)$ are provided in Fig.~\ref{fig:eff_vs_dalitz} to allow for the interpretation of the result in different theoretical model scenarios.

\begin{table}[b]
  \centering
  \caption{Efficiency ratios between the normalisation and signal modes and the single-event sensitivity, \Palpha, for the nominal \Xibc hypothesis, $m(\Xibc)=6.9\gevcc$ and $\tau(\Xibc)=400\fs$. The uncertainties are statistical only.}
\label{tab:alpha}  
  \begin{tabular}{|lccc|}
        \hline
        Period		&$\effCon/\effSig$	&\ncon		&\Palpha [$\times10^{-3}$] \\ \hline
        2016	&$3.66\pm0.17$	&$376\pm26$	&$9.7\pm0.8$	\\	 
        2017		&$3.50\pm0.13$	&$371\pm26$	&$9.4\pm0.7$ \\	 
        2018	&$3.22\pm0.13$	&$425\pm28$	&$7.6\pm0.6$ \\
        \hline
  \end{tabular}
\end{table}

\section{Systematic uncertainties}
\label{sec:sys}
Systematic uncertainties on the production ratio arise from the fit model, the trigger efficiency, the PID efficiency, the Dalitz plot weighting, and the simulation and data difference. The total systematic uncertainty is calculated as the quadratic sum of the individual uncertainties, presented in 
Table~\ref{tab:sys_summary},
assuming all the sources are uncorrelated.

The uncertainty on the signal yield may arise from the shape of the signal, the combinatorial background, and the misidentified background. This is quantified by choosing alternative functions. A Gaussian function is used for the signal and a second-order polynomial for the combinatorial background. The effect due to the misidentified background is estimated by fixing the ratio of the \LbDecaymis yield to that of the \LbDecay decay 
with their measured branching fractions~\cite{Aaij:2013pka}, taking into account their selection efficiencies. The sum in quadrature of these uncertainty estimates, yielding 3.6\%, is taken as systematic uncertainty due to the fit model.

The cancellation of the hardware-trigger efficiencies in the ratio between the signal and the normalisation modes is studied with \BDecay control samples, using a tag-and-probe method~\cite{Aaij:2012me}. The data and simulation difference between the efficiency ratio in the normalisation mode \LbDecay and the \BDecay control sample is assigned as a systematic uncertainty, and amounts to 6.8\%.

The PID efficiency is determined in bins of particle momentum and pseudorapidity using calibration data samples. There are several associated sources of systematic uncertainty, namely due to the limited size of the control samples, notably for high-\pt protons from the {\ensuremath{\PLambda}}\xspace sample, the assumption that kinematic correlations between tracks are neglected, and limitations in the method (\eg the finite kinematic binning used). The total systematic uncertainty associated with the PID efficiency, calculated as the sum in quadrature of individual contributions, amounts to 5.4\%.

The Dalitz plot of the simulated \LbDecay decays is weighted to match that observed in data. Several binning schemes of the Dalitz plot have been considered and the maximal difference in $R$ of 1.5\% is taken as the corresponding systematic uncertainty.

The simulation and data agreement is checked with control modes, and a difference of 5.0\% is found between different years of data-taking, which is taken as systematic uncertainty. 

 \begin{table}
  \centering
  \caption{Summary of the systematic uncertainties on measurement of the production ratio, $R$.}
  \label{tab:sys_summary}
  \begin{tabular}{|lr|}
    \hline
    Source              & $R$ [\%] \\
    \hline
    Fit model  & 3.6\\    
    Hardware trigger       & 6.8\\
    PID                 & 5.4\\
    \LbDecay Dalitz plot weight  & 1.5\\
    Simulation/data difference   & 5.0\\
    \hline
    Total               & 10.7\\
    \hline
    \end{tabular}
  \end{table}

\section{Variation of efficiency with mass and lifetime}
\label{sec:effmtau}
The trigger, reconstruction and selection efficiencies for \Xibc candidates have a strong dependence upon the \Xibc lifetime. The simulated \Xibc events are generated with a lifetime value of $400\fs$ as described in Sec.~\ref{sec:Detector}.
To test other lifetime hypotheses, the simulated events are weighted to reproduce other lifetime hypotheses and the efficiency is recalculated. A discrete set of hypotheses (100, 200, 300, 400 and $500\fs$) is considered. The total efficiency is found to have a linear dependence on the \Xibc lifetime. The value and uncertainty on the single-event sensitivity \Palpha are provided for each lifetime hypothesis in Table~\ref{tab:all_alpha_tau}. 

The efficiency could also depend on the \Xibc baryon mass hypothesis in the simulation, since it affects the kinematic distributions of the decay products. To assess its effect, large samples of simulated events are generated with alternative mass hypotheses, namely $6.7$ and $7.1\gevcc$. The efficiencies for other mass values are interpolated between the nominal and these two hypotheses. 
Two tests are carried out with these samples. Firstly, the detector acceptance efficiency is recomputed. Secondly, the \pt distributions of the \Xibc baryon daughters are weighted to match those of the alternative mass hypothesis and the remaining efficiency is recalculated. The total efficiency is found to have negligible dependence on the $\Xibc$ mass, thus it is ignored in the evaluation of the single-event sensitivities.

\begin{table}
  \centering
  \caption{Single-event sensitivity $\Palpha$ in units of $10^{-3}$ for different lifetime hypotheses of the \Xibc baryon for different data-taking periods. The uncertainties are due to the limited size of the simulated samples and the statistical uncertainties on the measured \Lb baryon yields.}
  \label{tab:all_alpha_tau}  
  \begin{tabular}{|lccccc|}
        \hline
    Period	&$\tau=100\fs$	&$\tau=200\fs$	&$\tau=300\fs$	&$\tau=400\fs$	&$\tau=500\fs$	\\ \hline
    2016		&$ 141 \pm 14$	&$ 27.5 \pm 2.4 $	&$ 14.1 \pm 1.2 $	&$ 9.7 \pm 0.8 $	&$ 7.7 \pm 0.7 $\\
   2017		&$ 134 \pm 12 $	&$ 25.9 \pm 2.1 $	&$ 13.5 \pm 1.1 $	&$ 9.5 \pm 0.8 $	&$ 7.6 \pm 0.6 $\\
   2018		&$ 102 \pm 9 $	&$ 20.8 \pm 1.6 $	&$ 10.8 \pm 0.8 $	&$ 7.6 \pm 0.6 $	&$ 6.1 \pm 0.5 $\\
    \hline
    \end{tabular}
\end{table}

\section{Results}
\label{sec:ul}
\begin{figure}
  \centering
  \includegraphics[width=0.75\linewidth]{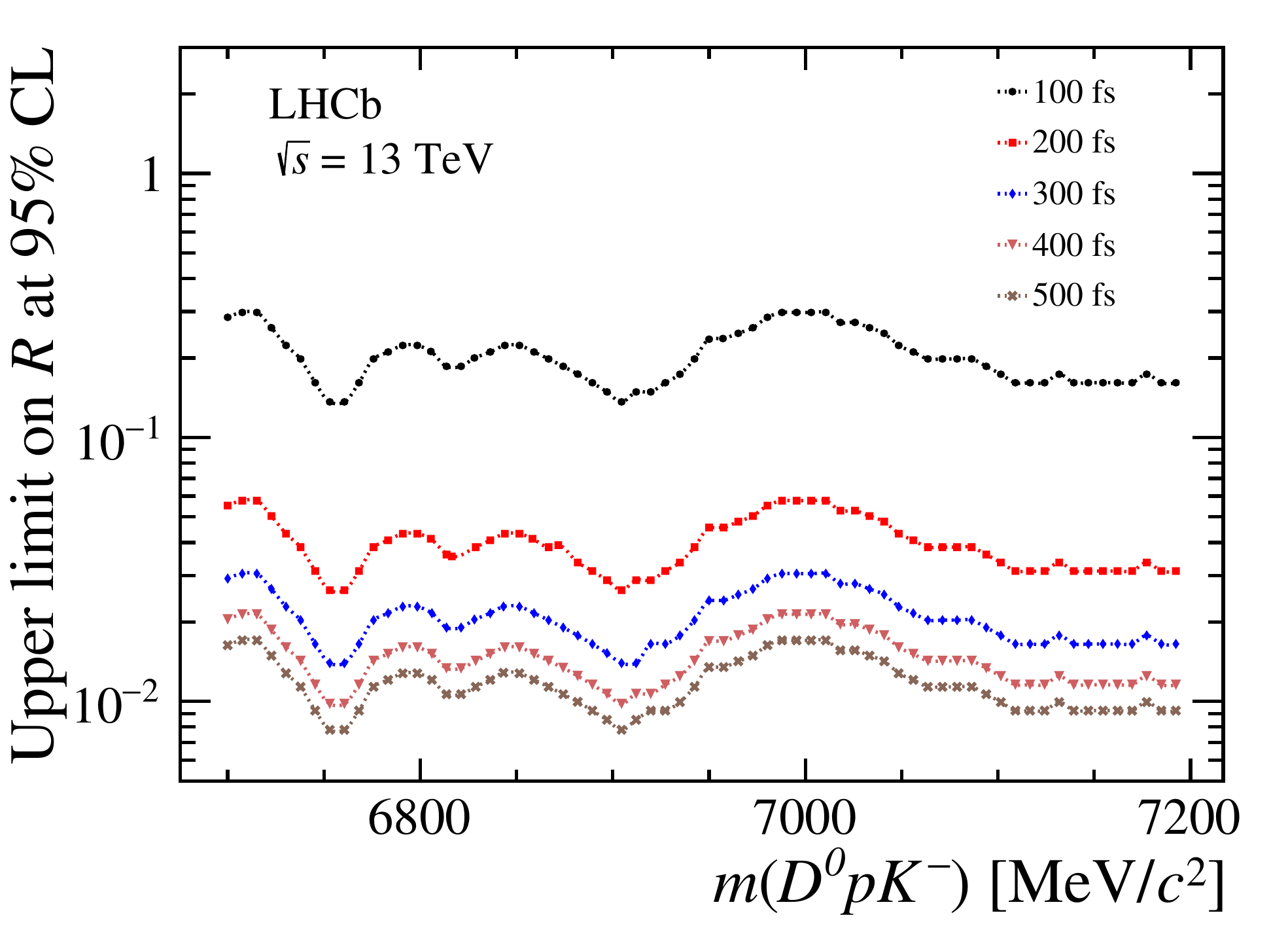}
  \caption{Values of upper limits on $R$ at 95\% CL as a function of $m(\Dz\proton\Km)$ for five $\Xibc$ lifetime hypotheses. The curves from top to bottom correspond to lifetime hypotheses from $100\fs$ to $500\fs$, respectively.}
  \label{fig:ul}
\end{figure}

The upper limits on the \Xibc decay ratio $R$ are obtained by performing again a fit to the data invariant mass distribution assuming different \Xibc mass hypotheses in the range from 6.7 to $7.2\gevcc$, and in steps of $7.5\mevcc$, for five lifetime hypotheses, in the fiducial region of rapidity
$2.0 < y < 4.5$ and transverse momentum $5 < \pt < 25 \gevc$. For each \Xibc baryon mass and lifetime hypothesis, the likelihood profile $\mathcal{L}(R)$ is determined as a function of $R$ with simultaneous fits to the $m(\Dz\proton\Km)$ invariant mass distributions. Then it is convoluted with a Gaussian distribution whose width is a quadratic sum of the statistical and systematic uncertainty on the single-event sensitivity. The upper limit at 95\% credibility level (CL) is defined as the value of $R$ at which the integral of the profile likelihood equals 95\% of its total area. Upper limits on $R$ at 95\% CL for different lifetime hypotheses are shown in Fig.~\ref{fig:ul}.

\section{Conclusion}
\label{sec:conclusion}
A first search for the $\XibcDecay$ decay is performed at LHCb with a data sample of $pp$ collisions, corresponding to an integrated luminosity of 5.4\invfb, recorded at a centre-of-mass energy of 13\tev. No evidence for a signal is found. Upper limits at 95\% CL on the \Xibc baryon production cross-section times its branching fraction to the $\Dz\proton\Km$ final state relative to the \LbDecay decay are obtained in the fiducial region of rapidity $2.0 < y < 4.5$ and transverse momentum $5 < \pt < 25 \gevc$, and for various \Xibc mass and lifetime hypotheses. The upper limits are set assuming that the kinematic distributions of the \Xibc baryon follow those of the \genxicctwo model~\cite{Chang:2009va} and that the decay of the \Xibc baryon proceeds according to a uniform phase-space model. The values of the upper limits depend strongly on the lifetime, varying from $3.0\times10^{-1}$ to $1.7\times10^{-2}$ for $100\fs$ and $500\fs$, respectively. Future searches at LHCb with improved trigger conditions, additional \Xibc decay modes, and larger data samples will further improve the \Xibc signal sensitivity.

%% file: acknowledgements.tex
\section*{Acknowledgements}
%
% These Acknowledgements valid from 3-May-2019
%
\noindent We thank Chao-Hsi Chang, Cai-Dian L\"u, Wei Wang, Xing-Gang Wu, and Fu-Sheng Yu for frequent and interesting discussions on the production and decays of double-heavy-flavor baryons.
We express our gratitude to our colleagues in the CERN
accelerator departments for the excellent performance of the LHC. We
thank the technical and administrative staff at the LHCb
institutes.
We acknowledge support from CERN and from the national agencies:
CAPES, CNPq, FAPERJ and FINEP (Brazil); 
MOST and NSFC (China); 
CNRS/IN2P3 (France); 
BMBF, DFG and MPG (Germany); 
INFN (Italy); 
NWO (Netherlands); 
MNiSW and NCN (Poland); 
MEN/IFA (Romania); 
MSHE (Russia); 
MICINN (Spain); 
SNSF and SER (Switzerland); 
NASU (Ukraine); 
STFC (United Kingdom); 
DOE NP and NSF (USA).
We acknowledge the computing resources that are provided by CERN, IN2P3
(France), KIT and DESY (Germany), INFN (Italy), SURF (Netherlands),
PIC (Spain), GridPP (United Kingdom), RRCKI and Yandex
LLC (Russia), CSCS (Switzerland), IFIN-HH (Romania), CBPF (Brazil),
PL-GRID (Poland) and OSC (USA).
We are indebted to the communities behind the multiple open-source
software packages on which we depend.
Individual groups or members have received support from
AvH Foundation (Germany);
EPLANET, Marie Sk\l{}odowska-Curie Actions and ERC (European Union);
A*MIDEX, ANR, Labex P2IO and OCEVU, and R\'{e}gion Auvergne-Rh\^{o}ne-Alpes (France);
Key Research Program of Frontier Sciences of CAS, CAS PIFI,
Thousand Talents Program, and Sci. \& Tech. Program of Guangzhou (China);
RFBR, RSF and Yandex LLC (Russia);
GVA, XuntaGal and GENCAT (Spain);
the Royal Society
and the Leverhulme Trust (United Kingdom).

%% file: LHCb_Authorship_12-May-2020.tex
% LHCb collaboration author list
% Data extracted on October 3rd, 2020 at 10:39am for reference date 12-May-2020
\centerline
{\large\bf LHCb collaboration}
\begin
{flushleft}
\small
R.~Aaij$^{31}$,
C.~Abell{\'a}n~Beteta$^{49}$,
T.~Ackernley$^{59}$,
B.~Adeva$^{45}$,
M.~Adinolfi$^{53}$,
H.~Afsharnia$^{9}$,
C.A.~Aidala$^{83}$,
S.~Aiola$^{25}$,
Z.~Ajaltouni$^{9}$,
S.~Akar$^{64}$
% ...

{\footnotesize \it

% ...
$ ^{9}$Universit{\'e} Clermont Auvergne, CNRS/IN2P3, LPC, Clermont-Ferrand, France\\
% ...
$ ^{25}$INFN Sezione di Milano, Milano, Italy\\
% ...
$ ^{31}$Nikhef National Institute for Subatomic Physics, Amsterdam, Netherlands\\
% ...
$ ^{45}$Instituto Galego de F{\'\i}sica de Altas Enerx{\'\i}as (IGFAE), Universidade de Santiago de Compostela, Santiago de Compostela, Spain\\
% ...
$ ^{49}$Physik-Institut, Universit{\"a}t Z{\"u}rich, Z{\"u}rich, Switzerland\\
% ...
$ ^{53}$H.H. Wills Physics Laboratory, University of Bristol, Bristol, United Kingdom\\
% ...
$ ^{59}$Oliver Lodge Laboratory, University of Liverpool, Liverpool, United Kingdom\\
% ...
$ ^{64}$University of Cincinnati, Cincinnati, OH, United States\\
% ...
$ ^{67}$Syracuse University, Syracuse, NY, United States\\
% ...
$ ^{83}$University of Michigan, Ann Arbor, United States, associated to $^{67}$\\
\bigskip
% ...
\medskip
}
\end{flushleft}